%% file: main.tex
\documentclass[runningheads,a4paper]{llncs}

\usepackage[T1]{fontenc}
\usepackage{lmodern}
\AtBeginDocument{\DeclareFontShape{T1}{lmr}{bx}{sc}{<->ssub*lmr/m/sc}{}}
\usepackage[utf8]{inputenc}
\usepackage{amsmath, amssymb}
\usepackage{mathtools}
\usepackage{booktabs}
\usepackage{graphicx}
\usepackage{xcolor}
\usepackage[hidelinks]{hyperref}
\usepackage{listings}
\lstdefinestyle{disepy}{
  language=Python,
  basicstyle=\ttfamily\small,
  keywordstyle=\bfseries,
  commentstyle=\itshape\color{gray},
  stringstyle=\color{black},
  showstringspaces=false,
  columns=fullflexible,
  keepspaces=true,
  aboveskip=0pt, belowskip=0pt,
  frame=none,
}
\newenvironment{dbox}%
  {\par\vskip4pt\noindent\hrule\vskip6pt}%
  {\par\vskip6pt\hrule\vskip4pt}
\makeatletter
\def\cref@split#1:#2\@nil{#1}
\newcommand{\cref@prefix}[1]{\expandafter\cref@split#1:\@nil}
\@namedef{cref@Name@sec}{Section}   \@namedef{cref@name@sec}{section}
\@namedef{cref@Name@app}{Appendix}  \@namedef{cref@name@app}{appendix}
\@namedef{cref@Name@def}{Definition}\@namedef{cref@name@def}{definition}
\@namedef{cref@Name@ass}{Assumption}\@namedef{cref@name@ass}{assumption}
\@namedef{cref@Name@lem}{Lemma}     \@namedef{cref@name@lem}{lemma}
\@namedef{cref@Name@thm}{Theorem}   \@namedef{cref@name@thm}{theorem}
\@namedef{cref@Name@cor}{Corollary} \@namedef{cref@name@cor}{corollary}
\@namedef{cref@Name@rem}{Remark}    \@namedef{cref@name@rem}{remark}
\@namedef{cref@Name@fig}{Figure}    \@namedef{cref@name@fig}{figure}
\@namedef{cref@Name@tab}{Table}     \@namedef{cref@name@tab}{table}
\@namedef{cref@Name@alg}{Algorithm} \@namedef{cref@name@alg}{algorithm}
\@namedef{cref@Name@eq}{Equation}   \@namedef{cref@name@eq}{equation}
\newcommand{\cref@word}[2]{%
  \@ifundefined{cref@#1@#2}{Section}{\@nameuse{cref@#1@#2}}}
\providecommand{\Cref}[1]{\cref@word{Name}{\cref@prefix{#1}}~\ref{#1}}
\providecommand{\cref}[1]{\cref@word{name}{\cref@prefix{#1}}~\ref{#1}}
\makeatother

\spnewtheorem*{assumption*}{Assumption}{\bfseries}{\itshape}

\newcommand{\inX}{\mathcal{X}}
\newcommand{\inY}{\mathcal{Y}}
\newcommand{\dist}{\mathcal{D}}
\newcommand{\prog}{P}
\newcommand{\prop}{\varphi}
\newcommand{\mutrue}{\mu}
\newcommand{\muhat}{\hat{\mu}}

\newcommand{\partset}{\Pi}
\newcommand{\openleaves}{\partset^{\mathrm{open}}}
\newcommand{\closedleaves}{\partset^{\mathrm{closed}}}
\newcommand{\Wopen}{W_{\mathrm{open}}}
\newcommand{\epsstat}{\varepsilon_{\mathrm{stat}}}
\newcommand{\hwhat}{\hat{w}}
\newcommand{\hpathat}{\hat{p}}

\newcommand{\unsat}{\textsf{unsat}}

\newcommand{\unknown}{\textsf{unknown}}
\newcommand{\toolname}{\textsc{DiSE}}
\newcommand{\code}[1]{\texttt{#1}}

\newcommand{\apx}[1]{the extended version~\cite{diseartifact}}
\newcommand{\Apx}[1]{The extended version~\cite{diseartifact}}

\begin{document}

\title{How Often Does Your Program Fail?}
\titlerunning{How Often Does Your Program Fail?}
\author{Arnab Ray\inst{1}\orcidID{0009-0000-8081-3981} \and Aalok Thakkar\inst{2}\orcidID{0000-0002-3195-585X}}
\authorrunning{A. Ray and A. Thakkar }
\institute{Indian Statistical Institute, Kolkata, India\\
           \email{rayarnab015@gmail.com}
           \and
           Ashoka University, Sonipat, India\\
           \email{thakkar@ashoka.edu.in}}


\maketitle

\begin{abstract}
Software in production encounters inputs shaped by how it is used in
practice. We study \emph{distribution-aware reliability estimation}:
given a program, its operational input distribution, and a condition
of interest, determine how often that condition holds and certify the
result with a guaranteed error bound.

Symbolic and statistical methods offer two ways to answer this
question. Symbolic methods reason about entire regions of the input
space and can certify rates exactly, but often struggle with complex
arithmetic or loops. Statistical methods instead sample inputs and
apply concentration bounds. They are broadly applicable, but can
require many samples when failures are rare. We bring these approaches
together in a framework that includes both as special cases. Each
estimator has three components: a mass estimator, a per-leaf confidence
bound, and a symbolic closure rule. We prove that any instantiation
satisfying three invariants returns an interval containing the true
rate with confidence $1-\delta$, regardless of when it stops. The
certified error has two parts: a statistical term, reduced by sampling,
and a structural term, reduced by symbolic closure. Pure sampling and
pure symbolic execution each reduce only one of these terms.
Alternative choices of the components also allow the framework to
support rare-event variance-reduction methods.

We present \textsc{DiSE}, an instantiation that combines an SMT-only
closure rule with a variance-adaptive empirical-Bernstein confidence
sequence. Its scheduler uses a single gain-per-cost criterion to
allocate effort between the two error terms, without hand-tuned
switching. We implement the framework as an open-source tool and
evaluate it on three benchmark families. In these experiments,
\textsc{DiSE} certifies rare-event reliabilities that pure sampling
cannot certify and resolves structurally concentrated conditions that
pure symbolic reasoning leaves unresolved. All instantiations retain
their soundness guarantees across the evaluated cases.
\end{abstract}


\input{macros}
\input{sections/introduction}
\input{sections/related}

\input{sections/problem}

\input{sections/DiSE}
\input{sections/implementation}
\input{sections/experiments}
\input{sections/conclusion}
\bibliographystyle{splncs04}
\bibliography{ref}

\end{document}

%% file: macros.tex
\newcommand{\domieq}{\succcurlyeq}
\newcommand{\size}[1]{\left| #1 \right|}
\newcommand{\floor}[1]{\left\lfloor #1 \right\rfloor}
\newcommand{\ceil}[1]{\left\lceil #1 \right\rceil}
\newcommand{\domi}{\succcurly}
\newcommand{\E}{\mathop{\mathbb{E}}}
\newcommand{\remove}[1]{}
\newcommand{\OPT}{\mathit{OPT}}
\newcommand{\nmax}{n_{\max}}
\newcommand{\R}{\mathbb{R}}
\newcommand{\N}{\mathbb{N}}
\newcommand{\cI}{\mathcal{I}}
\newcommand{\cJ}{\mathcal{J}}
\newcommand{\cM}{\mathcal{M}}
\newcommand{\cS}{\mathcal{S}}
\newcommand{\cT}{\mathcal{L}}
\newcommand{\cL}{\mathcal{L}}
\newcommand{\cA}{\mathcal{A}}
\newcommand{\cB}{\mathcal{B}}
\newcommand{\cC}{\mathcal{C}}
\newcommand{\cD}{\mathcal{D}}
\newcommand{\mD}{\mathrm{D}}
\newcommand{\cE}{\mathcal{E}}
\newcommand{\cF}{\mathcal{F}}
\newcommand{\cP}{\mathcal{P}}
\newcommand{\mP}{\mathrm{P}}
\newcommand{\cN}{\mathcal{N}}
\newcommand{\cK}{\mathcal{K}}
\newcommand{\cG}{\mathcal{G}}
\newcommand{\cO}{\mathcal{O}}
\newcommand{\cR}{\mathcal{R}}
\newcommand{\cW}{\mathcal{W}}
\newcommand{\Oh}{\mathcal{O}}
\newcommand{\cQ}{\mathcal{Q}}
\newcommand{\mQ}{\mathrm{Q}}
\newcommand{\tOh}{\widetilde{{\mathcal O}}}
\newcommand{\cCH}{\mathcal{CH}}
\newcommand{\cH}{\mathcal{H}}
\newcommand{\ch}{\mathcal{h}}
\newcommand{\cX}{\mathcal{X}}
\newcommand{\cY}{\mathcal{Y}}
\newcommand{\cU}{\mathcal{U}}
\newcommand{\mU}{\mathrm{U}}
\newcommand{\hs}{HS}
\newcommand{\mat}{{\sf Match}}
\newcommand{\pack}{{\sf Pack}}
\newcommand{\cut}{{\sf Cut}}
\newcommand{\eps}{\varepsilon}
\newcommand{\pr}{\mathbb{P}}
\newcommand{\ang}[1]{{\langle{#1}\rangle}}
\newcommand{\complain}[1]{\textcolor{red}{#1}}
\newcommand{\comments}[1]{\textcolor{blue}{\bf{#1}}}
\newcommand{\ind}{\mathbb{I}}

\newcommand{\spcl}{\rho}
\newcommand{\wt}{\texttt{wt}}
\newcommand{\dtv}{\mathsf{d_{TV}}}
\newcommand{\dkl}{\mathsf{d_{KL}}}
\newcommand{\dk}{\mathsf{d_{K}}}
\newcommand{\dchi}{\mathsf{d_{\chi^2}}}
\newcommand{\deltv}{\mathsf{\Delta_{TV}}}
\newcommand{\delkl}{\mathsf{\Delta_{KL}}}
\newcommand{\delchi}{\mathsf{\Delta_{\chi^2}}}
\newcommand{\uni}{\mathsf{Unif}}
\newcommand{\expo}{\mathsf{Exp}}
\newcommand{\pois}{\mathsf{Pois}}
\newcommand{\p}{\mathcal{P}}
\newcommand{\epstpa}{\eta}

\newcommand{\truedist}{\cD}
\newcommand{\truemarginal}[1]{\cD_{#1}}
\newcommand{\truedensity}{f_{\cD}}
\newcommand{\truemarginaldensity}[1]{f_{\cD_{#1}}}
\newcommand{\granulardiscrete}[1]{\cD_{#1}^{\cI}}
\newcommand{\discrete}[1]{\mD_{#1}^{\cI}}
\newcommand{\granularunifmarginal}[1]{\bar{\cD}_{#1}^{\cI}}
\newcommand{\hypothesis}{\cH}

\newcommand {\blfootnote}[1]{%
  \begingroup
  \renewcommand\thefootnote{}\footnote{#1}%
  \addtocounter{footnote}{-1}%
  \endgroup
}

\newtheorem{theo}{Theorem}[section]
\newtheorem{lem}[theo]{Lemma}
\newtheorem{pre}[theo]{Proposition}
\newtheorem{coro}[theo]{Corollary}
\newtheorem{algo}{Algorithm}
\newtheorem{conj}{Conjecture}
\newtheorem{cl}[theo]{Claim}
\newtheorem{defi}[theo]{Definition}
\newtheorem{rem}[theo]{Remark}
\newtheorem{assumption}{Assumption}

\newcommand{\prob}[1]{\mathrm{Pr}\left(#1\right)}
\newcommand{\samp}{\textsf{SAMP}}
\newcommand{\cond}{\textsf{COND}}

\setlength{\textfloatsep}{8pt plus 1.0pt minus 2.0pt}

\newcommand{\Ball}{\mathsf{Ball}}
\newcommand{\low}{\mathsf{low}}
\newcommand{\high}{\mathsf{high}}
\newcommand{\domain}{\Sigma}

\newcommand{\norm}[1]{\left\lVert #1 \right\rVert}
\newcommand{\Prf}{\Pr}

\newcommand{\Est}{\mathsf{Est}}

\newcommand{\mix}{\mathsf{Mix}}
\newcommand{\dx}{\mathrm{dx}}
\newcommand{\dy}{\mathrm{dy}}

\newcommand{\tpa}{\ensuremath{\mathsf{TPA}}\xspace}
\newcommand{\tpacore}{\ensuremath{\mathsf{TPACore}}\xspace}
\newcommand{\Thresh}{\ensuremath{\mathsf{Thresh}}\xspace}
\newcommand{\testlog}{\ensuremath{\mathsf{EQLogLipschitz}}\xspace}
\newcommand{\testlip}{\ensuremath{\mathsf{EQLipschitz}}\xspace}
\newcommand{\discsamp}{\ensuremath{\mathsf{DiscSamp}}\xspace}
\newcommand{\flatdec}{\ensuremath{\mathsf{ConstructFlat}}\xspace}
\newcommand{\birgelearner}{\ensuremath{\mathsf{Birge}}\xspace}
\newcommand{\dflat}{\mathcal D_{\mathsf{flat}}}
\newcommand{\dred}{\mathcal D_{\mathsf{red}}}
\newcommand{\mH}{\mathrm{H}}

%% file: sections/introduction.tex
\section{Introduction}\label{sec:intro}

Automated testing helps uncover bugs in safety-critical software,
including aircraft collision avoidance systems, autonomous vehicle
controllers, and medical device firmware. Finding a failure, however,
answers only part of the reliability question. We also need to know
how often the software fails under the inputs it encounters in
deployment. A counterexample shows that failure is possible; it does
not tell us whether the failure rate meets a required reliability
bound.

Consider a terrain avoidance warning system in a commercial aircraft.
Suppose automated testing identifies a combination of altitude,
descent rate, and terrain proximity that causes the warning to sound
too late for safe recovery. The fault is established. To assess its
operational significance, a certification authority also needs to
know how often those conditions occur across the fleet. Comparing
the failure rate with a numerical reliability threshold requires
more than a point estimate: it requires a bound on the uncertainty
in that estimate. The same issue arises whenever software operates
under a known or estimated input distribution and must meet a
contractual reliability requirement. An estimated failure rate of
$0.003$, for example, does not by itself tell us how confidently we
can conclude that a bound of $0.001$ has been exceeded.

The natural response to a known fault is to fix it. In practice,
though, the rate can matter both before and after a repair. It may
determine whether the component meets its reliability requirements
and whether a change is required. A third-party or separately
qualified component may be difficult to modify without reopening
certification. The behaviour of interest may fall within a tolerated
numerical or timing envelope rather than arise from a defect with a
local fix. Even when a candidate repair is available, its residual
failure rate may still need to be certified. In each case, the needed
result is a rate with an explicit confidence guarantee. Neither a
counterexample nor the failure to find one provides that result
(\Cref{sec:case-study}).

Two natural approaches address this problem, but their strengths and
limitations differ. The \emph{symbolic} approach reasons about program
paths and can compute the exact failure probability when the path
conditions and probability calculations are tractable. Complex
arithmetic or loops, however, can leave obligations that the reasoning
engine cannot resolve. The \emph{statistical} approach runs the
program on randomly sampled inputs and counts failures. It applies
even when symbolic reasoning is intractable, but rare failures make
it expensive: a failure affecting one input in ten thousand may
require many thousands of executions to observe even once, and many
more to estimate its rate precisely.

\paragraph{A unifying framework.}
We develop a framework that includes both approaches. Each estimator
is built from three interchangeable components: a mass estimator, a
per-leaf confidence bound, and a symbolic closure rule. We show that
any instantiation satisfying three invariants returns an interval
containing the true rate with confidence $1-\delta$, regardless of
when it stops (Theorem~\ref{thm:hw-decomp}).

The proof separates the certified error into two terms. The
statistical term $\epsstat$ captures the remaining uncertainty in
regions estimated by sampling. The structural term $\Wopen$ is the
mass of regions not yet resolved by either sampling or symbolic
closure. Sampling reduces the first term; symbolic closure reduces
the second. Pure sampling and pure symbolic execution are recovered
as the two \emph{degenerate} instantiations, each reducing only one
term. The framework also accommodates rare-event variance reduction
through alternative choices of its components
(\Cref{cor:certified-is}).

\paragraph{An adaptive instantiation.}
Our main instantiation, \textsc{DiSE}, works on both error terms.
It maintains a partition of the input space into regions induced by
the program's branching structure. At each step, a single
gain-per-cost criterion determines where to spend effort, rather
than relying on a hand-tuned rule for switching between symbolic
reasoning and sampling.

When the solver can certify a region, \textsc{DiSE} uses that result
without sampling it. When the solver cannot, \textsc{DiSE} samples
\emph{within that region}, where executions can be more informative
than samples drawn from the full input distribution. The error
decomposition thus serves two purposes: it establishes soundness
and guides the scheduler. The algorithm stops once the sum of the
two error terms falls below the target precision.

This combination is particularly useful when failures are rare but
concentrated in specific regions of the input space, a setting in
which both pure sampling and pure symbolic reasoning can struggle.
In our experiments on such programs, \textsc{DiSE} needs up to two
orders of magnitude fewer executions than plain sampling to certify
the same interval. We implement the framework in the open-source
tool \textsc{DiSE} and evaluate its instantiations across three
benchmark families (\Cref{sec:setup}). The evaluation shows how the
relative performance of different schedules depends on the program
class, while all instantiations preserve the soundness guarantee.

%% file: sections/related.tex
\section{Related Work}\label{sec:related}

\paragraph{Probabilistic symbolic execution and its optimizations.}
Geldenhuys, Dwyer, and Visser~\cite{geldenhuys2012pse} introduced
PSE, combining symbolic execution with model counting to compute
exact path probabilities for affine path conditions over uniform
inputs.  Subsequent work has focused on reducing path-enumeration and model-counting costs.  On the counting
side, Filieri et al.~\cite{filieri2013spf} widen the
countable fragment beyond uniform integer domains and integrate model
counting into Symbolic PathFinder; Borges et
al.~\cite{borges2014compositional} quantify solution spaces
\emph{compositionally}, reusing counts across path conditions sharing
structure, and later replace exact counting on intractable constraints
with an iterative distribution-aware
sampler~\cite{borges2015iterative}; Chistikov et
al.~\cite{chistikov2015approximate} give approximate counting directly
in SMT with PAC guarantees.  On the path side, Sankaranarayanan et
al.~\cite{sankaranarayanan2013static} infer whole-program bounds from
\emph{finitely many} paths, bounding the residual rather than
enumerating it, and Luckow et al.~\cite{luckow2014exact} extend PSE to
nondeterministic programs.  For an overview, see~\cite{pasareanu2022lecture}.

Two features separate \toolname{} from this line.  First the
deliverable: PSE and its optimizations return an exact count, or an
interval derived from one, and are restricted to the path conditions
their counter handles, whereas \toolname{} returns a
$(1-\delta)$-coverage interval and is therefore defined on
\emph{every} program, degrading to a wide but sound interval where PSE
returns nothing, the case for unbounded loops, nonlinear arithmetic,
and distributions that do not factor over the branch predicates.
Second the residual: where~\cite{sankaranarayanan2013static} bounds the
unenumerated mass and stops, \toolname{} \emph{samples inside} it,
converting the structural term $\Wopen$ into a statistical term
$\epsstat$ that further work shrinks.  When every region admits a
closed-form mass and closure succeeds on every leaf,
$\epsstat = \Wopen = 0$ and the two agree exactly; \code{pse\_exact}
in \Cref{sec:results} is that degenerate schedule.

\paragraph{Rare-event simulation and variance reduction.}
When the target event is rare, plain Monte Carlo is replaced by an
estimator with lower variance: importance
sampling~\cite{lecuyer2009importance,rubino2009rareevent}, which draws
from a tilted proposal and reweights; the cross-entropy
method~\cite{rubinstein2004crossentropy,jegourel2012crossentropy},
which learns that proposal; and multilevel
splitting~\cite{kahn1951splitting,jegourel2013splitting}, which
decomposes a rare event into a chain of likelier ones.  Statistical
model checking has adopted all
three~\cite{legay2010smc,jegourel2012crossentropy,jegourel2013splitting}.

These techniques can be incorporated into the framework’s estimators:  they
reduce the variance of an estimate of a fixed quantity.  Our
contribution is the \emph{schedule} over symbolic and statistical
actions, together with the soundness contract of
Theorem~\ref{thm:hw-decomp}, which is agnostic to how any individual
leaf is estimated.  Concretely, such an estimator is not a rival of
the framework but an instantiation of it: it occupies the
mass-estimator slot (I1) or the per-leaf-bound slot (I3), and any
instantiation meeting those invariants inherits $(1-\delta)$-coverage
(\Cref{sec:rare-event-instantiation}).  \toolname{}'s
importance-sampling fallback for non-axis-aligned regions is one such
instantiation.  The distinction that matters for evaluation is the
\emph{deliverable}: these methods typically report a point estimate or
an asymptotic confidence interval, whereas the contract of
\Cref{sec:problem} demands a finite-sample interval valid at a
data-dependent stopping time.  A head-to-head comparison must
therefore fix the deliverable and measure samples to a \emph{certified}
interval.

\paragraph{Statistical and probabilistic model checking.}
PRISM~\cite{kwiatkowska2011prism}, Storm~\cite{hensel2022storm} and
their statistical
variants~\cite{boyer2013plasma,hartmanns2014modest,schoolderman2024storm}
verify quantitative temporal properties on Markov chains and MDPs via
dedicated modelling languages.  \toolname{} operates directly on
source, treating the program as the model; the trade-off is
complementary, since \toolname{} targets Boolean output properties of a
deterministic program (\Cref{def:bool-property}) and does not resolve
nondeterminism, while PMC tools verify temporal specifications but
require a hand-built model for source-level loops.

\paragraph{Model counting.}
Approximate model counting~\cite{chakraborty2013approxmc,yang2023approxmc6}
gives PAC-$(\varepsilon,\delta)$ counts for propositional formulae,
with recent work adding formal certificates~\cite{kiesl2024certified}.
Delegating region-mass estimation to a certified counter is the
natural way to discharge \Cref{cor:certified-is} with tighter
constants than importance sampling.

\paragraph{Anytime-valid concentration inequalities.}
Hoeffding and Wilson bounds are valid only at a fixed sample size.
Howard et al.~\cite{howard2021time} give time-uniform bounds via
non-negative supermartingales; Waudby-Smith and
Ramdas~\cite{waudbysmith2024betting} sharpen this with the closed-form
PrPl-EB confidence sequence, variance-adaptive and matching Bennett's
inequality without known variance; Ramdas et
al.~\cite{ramdas2023gametheoretic} survey the area.  \toolname{} uses
PrPl-EB per leaf with Bonferroni correction over the bounded leaf
count.

%% file: sections/problem.tex
\section{Problem Formulation}\label{sec:problem}

We fix a finite input domain $\inX \subseteq \mathbb{Z}^d$ and assume
throughout:

\begin{assumption}\label{ass:program}
  The program $\prog : \inX \to \inY$ is deterministic over a finite input domain $\cX \subseteq \mathbb{Z}^d$. Randomized programs are modeled by treating the random seed as an additional input.
\end{assumption}

\noindent We certify how $\prog$ behaves on inputs drawn from a distribution
$\cD$, which requires fixing the class of input distributions and the
notion of a Boolean output property.

\begin{definition}[Discrete Product Distribution (DPD)]
\label{def:d1}
A probability distribution $\mathcal D:\mathcal X\rightarrow[0,1]$ over
$\mathcal X\subseteq\mathbb Z^d$ is a \emph{discrete product distribution}
if it factorizes as
\[
\mathcal D(x_1,\ldots,x_d)=\prod_{i=1}^{d}\mathcal D_i(x_i),
\]
where $\mathcal D_i$ is the marginal distribution of the $i$-th coordinate.
We assume that each marginal admits a closed-form cumulative probability mass
function, so the probability of any integer interval satisfies
\[
\mathcal D_i(\{a,\ldots,b\})=\mathcal D_i(b)-\mathcal D_i(a-1).
\]
\end{definition}

\begin{definition}[Bounded geometric distribution]
\label{def:bg}
Let $p\in(0,1)$ and $N\in\mathbb N$. The \emph{bounded geometric distribution}
$\mathrm{BG}(p,N)$ is the geometric distribution truncated to the support
$\{0,\ldots,N-1\}$. Its probability mass function is
\[
\mathcal D(k)=\frac{p(1-p)^k}{1-(1-p)^N},
\qquad k\in\{0,\ldots,N-1\},
\]
and its tail probability is
\[
\Pr[k\ge t]
=\frac{(1-p)^t-(1-p)^N}{1-(1-p)^N},
\qquad 0\le t<N.
\]
The product distribution $\mathrm{BG}(p,N)^{\otimes d}$, obtained by drawing
each coordinate independently from $\mathrm{BG}(p,N)$, is a discrete product
distribution (Definition~\ref{def:d1}).
\end{definition}

\begin{definition}[Output Property]
\label{def:bool-property}
An \emph{output property} for a deterministic program $P: \cX \to \cY$ is a measurable Boolean function over the output range $\varphi : \cY \to \{0,1\}$.
\end{definition}

Given a deterministic program $P$, a distribution $\mathcal{D}$
over its inputs reflecting operational usage, and a Boolean
property $\varphi$ encoding the condition of interest, we wish
to certify how often $P$ satisfies $\varphi$ under $\mathcal{D}$.
The central quantity is the \emph{operational reliability}.

\begin{definition}[Operational Reliability and Failure Set]
\label{def:reliability}
Let $P : \mathcal{X} \to \mathcal{Y}$ be a deterministic program,
$\mathcal{D}$ a discrete product distribution over $\mathcal{X}$,
and $\varphi : \mathcal{Y} \to \{0,1\}$ a Boolean output property.
The \emph{operational reliability} of $P$ under $\mathcal{D}$ with
respect to $\varphi$ is
\[
  \mu \;=\; \Pr_{x \sim \mathcal{D}}\bigl[\varphi(P(x)) = 1\bigr]
       \;=\; \sum_{x \in \mathcal{X}} \mathcal{D}(x)\,\varphi(P(x)).
\]
The \emph{failure set} of $P$ with respect to $\varphi$ is
$\mathcal{F}(P,\varphi) = \{\, x \in \mathcal{X} \mid
\varphi(P(x)) = 0 \,\}$. Since $\mathcal{D}(\mathcal{X}) = 1$,
it follows that $\mu = 1 - \mathcal{D}(\mathcal{F}(P,\varphi))$.
\end{definition}

\noindent Standard verification asks $\forall x : \varphi(P(x))$, that
is, whether $\mathcal{F}(P,\varphi)$ is empty; reliability estimation
asks for its \emph{measure} under $\cD$.  A counterexample certifies
non-emptiness and says nothing about $\mu$.

\begin{dbox}
\textbf{Reliability Estimation Problem.}
Given a deterministic program $P:\mathcal X\rightarrow\mathcal Y$, a
discrete product distribution $\mathcal D$ over $\mathcal X$, a Boolean
output property $\varphi:\mathcal Y\rightarrow\{0,1\}$, an error tolerance
$\varepsilon\in(0,1]$, and a confidence parameter $\delta\in(0,1)$,
compute an estimate $\hat{\mu}$ of the operational reliability $\mu$
together with an interval $[L,U]\subseteq[0,1]$ such that
\[
\Pr\!\left[L\le\mu\le U\right]\ge 1-\delta,
\qquad
U-L\le 2\varepsilon.
\]
The probability is taken over the algorithm's internal randomness and the
iid samples drawn from $\mathcal D$. The objective is to satisfy this
guarantee using as few evaluations of $P$ as possible.
\end{dbox}

\noindent The contract couples two requirements that are each trivial
alone: a wide interval achieves coverage, a point estimate achieves
width.  Attaining both at once, in as few evaluations as possible, is
the problem.

\paragraph{Semantic assumptions.}
Our implementation assumes deterministic programs over a finite domain
$\mathcal X\subseteq\mathbb Z^d$, discrete product input distributions,
Boolean output properties, and linear integer arithmetic. Inputs are interpreted
as mathematical integers rather than fixed-width machine words; this semantics is
exact for our benchmarks because every input range is bounded and represented
without overflow. Programs whose behavior depends on wraparound or bit-level
operations fall outside the supported fragment. During symbolic execution, traces
containing such operations are marked as concretized and remain open, contributing
only to $\Wopen$ and therefore affecting interval width but not soundness.

These assumptions are specific to the implementation of \toolname{}, not to the
framework itself. Theorem~\ref{thm:hw-decomp} relies only on the invariants of
Definition~\ref{def:invariants}; hence any implementation satisfying
(I1)-(I3) inherits the same $(1-\delta)$ coverage guarantee. In our
implementation, determinism allows a single \unsat{} result to certify the value
of $\varphi$ on a closed leaf (\Cref{lem:closure-soundness}), while the product
structure enables exact computation of region masses from marginal CDFs, yielding
$\beta=0$ (\Cref{lem:mass-conservation}). Other semantic models, such as
bitvectors, require different implementation components but do not change the
soundness argument.

\input{sections/illustrative_ex}

%% file: sections/illustrative_ex.tex
\subsection{Illustrative Example}
\label{sec:example}

\begin{figure}[ht]
\begin{center}
\begin{lstlisting}
        def monitor(x: int) -> int:
            if x < 50:          return SAFE
            if condition(x):    return FAULT
            return SAFE
\end{lstlisting}
\end{center}
\caption{Program \texttt{monitor}.  Under
$\cD = \mathrm{Geometric}(p=0.1)$ the region $\texttt{x >= 50}$ carries
mass $\Pr[x \ge 50] = 0.9^{49} \approx 0.0057$, so
$\mu \ge 0.9943$: the fault can affect at most $0.6\%$ of production
inputs.}
\label{fig:monitor}
\end{figure}

Consider \texttt{monitor} (\Cref{fig:monitor}), a simplified safety
monitor whose inputs are geometrically distributed, small values being
exponentially more common.  Its first branch covers almost all
production traffic and always returns \texttt{SAFE}; the fault can fire
only on inputs at least $50$, and only when \texttt{condition(x)} holds.
Which method fails depends entirely on that condition, and the same
program defeats each of the two for opposite reasons.

\emph{Symbolic methods fail on the condition.}  If
\texttt{condition(x)} lies outside the solver's decidable fragment, a
quadratic congruence say, the solver returns \unknown{} and the branch
is never resolved.  Probabilistic symbolic execution fails at the same
point for a sharper reason: it must \emph{count} the path condition's
satisfying inputs, and counting integer solutions of a quadratic
congruence needs algebraic number theory or enumeration.

\emph{Statistical methods fail on the distribution.}
\texttt{condition(x)} is reachable only when $x \ge 50$, a region of
mass $0.0057$: roughly one draw in $176$ lands there at all, and the
rest observe \texttt{SAFE} and say nothing about whether the fault
fires.  Certifying $\mu$ to half-width $\varepsilon$ costs
$\Theta(1/\varepsilon^{2})$ samples, and at the precisions a
reliability threshold demands that budget is infeasible.

\paragraph{How the algorithm proceeds.}
\toolname{} tracks the mass not yet resolved by either method and the
statistical uncertainty over what has been sampled but not certified,
stopping once their sum falls below $\varepsilon$.  On \texttt{monitor}
a small bootstrap sample lands almost entirely below $50$, all
traversing the clause $x < 50$; one solver call establishes that no
input satisfying $x < 50$ takes a different path, certifying the whole
region, whose mass $0.9943$ is read off the geometric CDF at zero
sampling cost.  One call has thus resolved $99.43\%$ of the mass and
driven the unresolved term to $0.0057$.  On the residual
$\{x \ge 50\}$ the solver returns \unknown{}, so \toolname{} asks
instead whether the residual matters; at $\varepsilon = 10^{-3}$ it
does, and the algorithm samples \emph{inside that region only}, where
each draw is $176$ times more informative than a draw from $\cD$ at
large, until the two terms sum below $\varepsilon$.  Neither method has
to work everywhere, and the stopping rule accounts for both sources of
uncertainty at once.

%% file: sections/DiSE.tex
\section{The Framework and the \toolname{} Algorithm}\label{sec:algorithm}

We now formalise the framework and the algorithm, fixing a
deterministic program $\prog \!:\! \inX \!\to\! \inY$, a discrete
product distribution $\cD$, a Boolean output property
$\prop \!:\! \inY \!\to\! \{0,1\}$, and parameters
$\varepsilon, \delta \in (0,1)$; the objective is the soundness
contract of \Cref{sec:problem}.  \Cref{sec:frontier} defines the
frontier and the estimator it induces, \Cref{sec:refine-close} the two
operations on it and the three invariants, \Cref{sec:hw-decomp} the
decomposition theorem, \Cref{sec:rare-event-instantiation} what may
occupy the component slots, and \Cref{sec:asip-loop} the driver loop.
Proofs are in the extended version.

\subsection{The Frontier}\label{sec:frontier}

A \emph{region} is a subset of $\inX$ defined by a quantifier-free
formula $F$ over the input variables in linear integer arithmetic
with equality and propositional connectives.
A \emph{frontier} $\partset$ is a finite tree of regions whose
leaves partition $\inX$; regions where $\prog$ behaves uniformly
are resolved exactly via closure, while uncertain regions are
handled statistically.
Each leaf $\pi$ carries:
\begin{itemize}
\item a region formula $F_\pi$, defining
  the region $R_\pi = \{x \in \inX : x \models F_\pi\}$;
\item a mass estimate $\hwhat_\pi$;
\item a finite observation sequence
  $\{(\mathrm{path}_i, \varphi_i)\}_{i=1}^{n_\pi}$, each entry
  the branch sequence and $\prop$-value of one concrete run of
  $\prog$ on an iid sample from $\dist_{\mid R_\pi}$;
\item a hit-rate estimate $\hpathat_\pi$ of the quantity $\Pr[\prop(\prog(X))=1 \mid X \in R_\pi]$;
\item a \emph{status} in
  $\{\mathsf{OPEN},\mathsf{CLOSED\text{-}T},
   \mathsf{CLOSED\text{-}F},\mathsf{EMPTY}\}$.
\end{itemize}
A leaf is $\mathsf{OPEN}$ until the oracle certifies $p_\pi$ to be
exactly $0$ or $1$ ($\mathsf{CLOSED\text{-}F}$,
$\mathsf{CLOSED\text{-}T}$) or finds $F_\pi$ unsatisfiable
($\mathsf{EMPTY}$), and the frontier starts as one $\mathsf{OPEN}$ leaf
$\pi_0$ with $F_{\pi_0} = \top$, $\hwhat_{\pi_0} = 1$.  Writing
$\openleaves$ and $\closedleaves$ for the $\mathsf{OPEN}$ and
$\mathsf{CLOSED\text{-}T}$ leaves, the induced estimator is
\begin{equation}\label{eq:muhat-def}
  \muhat \;=\;
    \sum_{\pi \in \closedleaves} \hwhat_\pi
    \;+\;
    \sum_{\pi \in \openleaves} \hwhat_\pi \cdot \hpathat_\pi,
\end{equation}
$\mathsf{CLOSED\text{-}F}$ and $\mathsf{EMPTY}$ leaves contributing
nothing; on an open leaf with no samples $\hpathat_\pi$ is an
arbitrary default in $[0,1]$, which does not affect soundness.  The
two sums are the two sources of uncertainty.

\subsection{Refinement, Closure, and Three Invariants}\label{sec:refine-close}

Two operations act on the frontier, and both are governed by three
pluggable components, a mass estimator, a closure rule and a per-leaf
bound, whose correctness conditions are the invariants below;
soundness reduces entirely to these three (\Cref{sec:hw-decomp}).

\paragraph{The two operations.}
Refinement of $\pi$ along a clause $b$ replaces it with children with
region formulas $(F_\pi \land b)$ and $(F_\pi \land \lnot b)$; since
$b$ is harvested from $\pi$'s own observed paths, refinement never
introduces predicates the program does not branch on.  Closure marks a
leaf \textsc{closed-t} or \textsc{closed-f} once $\prop(\prog(x))$ is
certified constant on all of $R_\pi$.  The mass estimator assigns
children their masses; the closure rule decides when to close; each
sampled leaf carries a per-leaf bound on its hit rate.

\begin{definition}[The three invariants]\label{def:invariants}
  Fix target confidence $(1-\delta)$ and an a priori cap
  $K_{\max}$ on the number of leaves the algorithm may create.
  \begin{description}
  \item[\textnormal{(I1) Mass conservation with budget $\beta$.}]
    Leaf masses sum to one ($\sum_\pi \hwhat_\pi = 1$), refinement
    preserves the sum, and the total mass error satisfies
    $\sum_\pi |\hwhat_\pi - w_\pi| \le \beta$.
  \item[\textnormal{(I2) Closure soundness.}]
    If the closure rule closes $\pi$ on value $v$, then
    $\prop(\prog(x))=v$ for every $x \in R_\pi$.
  \item[\textnormal{(I3) Anytime validity.}]
    Write $\hpathat_\pi(n)$ for the running hit-rate estimate at leaf
    $\pi$ after its first $n$ observations, that is the fraction of
    those $n$ runs with $\varphi$-value $1$, and $W_\pi(n)$ for the
    half-width the per-leaf bound reports at that sample count; both
    are random, and $W_\pi(n)$ is non-increasing in $n$.  The bound is
    \emph{anytime-valid} at confidence $1-\delta_\pi$,
    $\delta_\pi = \delta/K_{\max}$, if
    $\Pr[\,\exists n: |\hpathat_\pi(n)-p_\pi|>W_\pi(n)\,]
    \le \delta_\pi$,
    holding simultaneously at every sample count and hence at any
    data-dependent stopping time.
  \end{description}
\end{definition}

The decomposition theorem depends on nothing beyond these three
conditions.  $K_{\max}$ is fixed before the run and enforced by the
driver loop via a refinement-depth bound, so the Bonferroni allocation
$\delta_\pi = \delta/K_{\max}$ is \emph{a priori} and the union bound
behind Theorem~\ref{thm:hw-decomp} is valid however many leaves the run
creates.  A larger $K_{\max}$ widens each per-leaf bound by a
$\log K_{\max}$ factor but never threatens soundness.

\paragraph{Why anytime validity is needed.}
A classical interval, Hoeffding's or Wilson's, is a promise about
\emph{one} pre-declared sample size: fix $n$ before looking at the
data, and the resulting interval covers the truth with probability
$1-\delta$.  The promise lapses as soon as $n$ is chosen by looking at
the data, which is what any adaptive schedule does, ours included, and
stopping when an interval first looks satisfactory inflates the true
error rate~\cite{robbins1970law}.  A \emph{confidence sequence} makes
the stronger promise of an infinite family of intervals, one per sample
count, such that with probability $1-\delta$ \emph{every} member covers
the truth at once~\cite{howard2021time,ramdas2023gametheoretic}; being
simultaneous over $n$, it survives any stopping rule whatsoever. The coverage guarantee therefore remains valid under adaptive scheduling.

The next two lemmas discharge invariants (I1) and (I2) for
\toolname{}'s components; an instantiation may discharge them by any
other means.

\begin{lemma}[Exact mass on a bounded domain]%
\label{lem:mass-conservation}
  On the finite domain $\inX$, the mass of a region is the exact
  sum $w_\pi = \sum_{x \in R_\pi}\dist(x)$.  The estimator
  returning this sum in closed form as a product of
  marginal-CDF differences when $R_\pi$ is an axis-aligned box,
  and by enumeration of the enclosing box otherwise, satisfies
  (I1) with $\beta = 0$.
\end{lemma}

\noindent Every benchmark domain in \Cref{sec:results} is small enough
to enumerate, so $\hwhat_\pi = w_\pi$ and $\beta = 0$ throughout; the
budget $\beta$ is carried through the framework for the estimated-mass
instantiations of \Cref{sec:rare-event-instantiation}.

\toolname{} certifies closure of a leaf node via the SMT oracle.
Let $\mathrm{path}_\pi$ be the conjunction of branch clauses
recorded by the algorithm.

\begin{lemma}[Closure rule satisfies (I2)]%
\label{lem:closure-soundness}
  Suppose (i) the SMT oracle returns $\unsat{}$ for
  $F_\pi \wedge \neg\,\mathrm{path}_\pi$, and (ii) every
  observation at $\pi$ shares the same branch sequence
  $\mathrm{path}_\pi$ and $\varphi$-value $v$, and (iii) the SMT oracle
  proves that every execution following $\mathrm{path}_\pi$ has
  property value $v$.  Then
  $\prop(\prog(x)) = v$ for every $x \in R_\pi$.
\end{lemma}

\noindent Premise~(i) establishes that all inputs in $R_\pi$ follow
$\mathrm{path}_\pi$.  Premise~(iii) establishes that this path has a
constant property value $v$.  Premise~(ii) is a constant-time scan used
as a pre-filter before invoking the SMT query; if observations disagree,
the leaf remains \textsc{open}.  Closure therefore depends only on the
SMT certificates, reducing (I2) to the oracle's soundness.

\subsection{The Central Decomposition Theorem}\label{sec:hw-decomp}

An open leaf is \emph{sampled} if it carries at least one observation
($n_\pi \ge 1$) and \emph{unresolved} if it carries none, having been
touched by neither sampling nor closure.  The certified half-width has
one term per kind.

\begin{definition}[Half-width components]\label{def:hw-components}
  Let $W_\pi(T)$ be a per-leaf bound satisfying (I3)
  (\Cref{def:invariants}).  The \emph{statistical} half-width
  sums over the sampled open leaves, and the
  \emph{unresolved-mass} half-width over the unresolved ones:
  \begin{equation}
    \epsstat(T) = \!\!\sum_{\pi\ \mathrm{open,\ sampled}}\!\!
      \hwhat_\pi\,W_\pi(T),
    \qquad
    \Wopen(T)   = \!\!\sum_{\pi\ \mathrm{open,\ unresolved}}\!\!
      \hwhat_\pi .
  \end{equation}
\end{definition}

A \textsc{Sample} on an unresolved leaf moves its mass out of $\Wopen$
into $\epsstat$, where further samples shrink it; a successful
\textsc{Refine} closes a leaf and removes its mass from $\Wopen$
outright.  Symbolic certification is therefore the only way to drive a
leaf's contribution to exactly zero.

\begin{theorem}[Decomposition]\label{thm:hw-decomp}
  Consider an estimator built from a mass estimator, a closure
  rule, and a per-leaf bound satisfying invariants (I1), (I2) and
  (I3) of \Cref{def:invariants} with mass budget $\beta$.  Under
  Assumption~\ref{ass:program} and soundness of the SMT oracle, the
  estimator $\muhat_T$ of~\eqref{eq:muhat-def} satisfies, at
  every data-dependent stopping time $T$,
  \begin{equation}\label{eq:hw-decomp-formal}
    \Pr\!\Bigl[\,\bigl|\muhat_T - \mutrue\bigr|
      \;\le\; \epsstat(T) + \Wopen(T) + \beta\,\Bigr]
    \;\ge\; 1 - \delta .
  \end{equation}
\end{theorem}

\noindent\emph{Intuition.}  The error splits three ways.  Closed
leaves contribute none: (I2) fixes their $\varphi$-value exactly.
Sampled leaves contribute their statistical error, each within $W_\pi$
of $p_\pi$ by (I3), with a Bonferroni union bound over the
$\le K_{\max}$ leaves capping the total at $\epsstat$ and anytime
validity keeping this sound at the data-dependent $T$.  Unresolved
leaves have an unknown hit rate and are bounded by their own mass,
summing to $\Wopen$.  Mass error contributes at most $\beta$ by (I1).

\noindent The hypotheses name three pluggable components and nothing
else, so any estimator supplying them inherits $(1-\delta)$-coverage
at the data-dependent stopping time; the schedule governs only how
fast $\epsstat$, $\Wopen$ and $\beta$ fall.  The reported interval is
the clipped projection
\begin{equation}\label{eq:interval}
  [\hat L, \hat U] =
  \bigl[\max(0,\,\muhat_T - h),\;
        \min(1,\,\muhat_T + h)\bigr],
  \qquad h = \epsstat + \Wopen + \beta .
\end{equation}

\subsection{Instantiating the Estimator Slots}%
\label{sec:rare-event-instantiation}

Because Theorem~\ref{thm:hw-decomp} constrains the three components
only through (I1)--(I3), the slots they occupy are open to any
estimator meeting them, and in particular to the variance-reduction
machinery developed for rare-event simulation
(\Cref{sec:related}).  The following corollary formalises this integration.

\toolname{} discharges (I1) with $\beta = 0$ by enumeration
(\Cref{lem:mass-conservation}), which is available only because the
domain is bounded and $\cD$ factors.  Where it is not, the mass of a
leaf must itself be estimated, and importance sampling is the standard
device: draw from a proposal $q$ supported on $R_\pi$ and reweight by
$\cD/q$.  A cross-entropy
scheme~\cite{rubinstein2004crossentropy,jegourel2012crossentropy}
tunes $q$; multilevel splitting~\cite{jegourel2013splitting} replaces
the single estimate by a telescoping product.  What the framework
requires of any of them is not a point estimate but a
\emph{certificate} on that estimate.

\begin{corollary}[Certified estimator slots]\label{cor:certified-is}
  Split the confidence budget as $\delta = \delta_{\mathrm{stat}} +
  \delta_{\mathrm{mass}}$.  Suppose the closure rule satisfies (I2);
  the per-leaf bound satisfies (I3) at confidence
  $1-\delta_{\mathrm{stat}}/K_{\max}$; and the mass estimator returns,
  for each leaf $\pi$, an estimate $\hwhat_\pi$ with
  $\sum_\pi \hwhat_\pi = 1$ together with a bound $b_\pi$ that is
  anytime-valid at confidence $1-\delta_{\mathrm{mass}}/K_{\max}$,
  \[
    \Pr\bigl[\,\exists m :\;
      |\hwhat_\pi(m) - w_\pi| > b_\pi(m)\,\bigr]
      \;\le\; \delta_{\mathrm{mass}}/K_{\max},
  \]
  where $m$ is the number of draws the estimator has spent on $\pi$.
  Then at every data-dependent stopping time $T$,
  \[
    \Pr\Bigl[\,\bigl|\muhat_T - \mutrue\bigr| \;\le\;
      \epsstat(T) + \Wopen(T) + \textstyle\sum_{\pi} b_\pi(T)
      \,\Bigr] \;\ge\; 1-\delta .
  \]
\end{corollary}

\noindent The proof is a union bound over the two families of events
followed by Theorem~\ref{thm:hw-decomp} applied on their intersection
with $\beta = \sum_\pi b_\pi(T)$; it is given in \apx{app:certified-is}.  Three consequences are worth stating.

The choice of variance-reduction method affects efficiency; coverage follows from the certificate.  A
better proposal $q$ shrinks $b_\pi$ at a fixed budget and so tightens
the reported interval, but it cannot make the interval unsound,
because soundness passes through the certificate and not through the
estimator.  Cross-entropy tuning, splitting, and plain rejection
sampling are in this sense interchangeable: they are different points
on the cost-versus-$b_\pi$ curve of the same slot.

Second, the certificate must be \emph{anytime-valid} for the same
reason the per-leaf bound must be.  A schedule that decides how many
importance samples to spend on a leaf by watching the estimate settle
is peeking, and a fixed-$m$ interval, an asymptotic normal
approximation or a bootstrap interval among them, does not survive it.
This is the technical sense in which the rare-event literature's usual
deliverable, a point estimate or an asymptotic confidence interval,
does not by itself discharge (I1): it must first be paired with a
finite-sample, time-uniform bound.  Empirical-Bernstein confidence
sequences~\cite{waudbysmith2024betting,howard2021time} supply one
whenever the weights $\cD/q$ are bounded, which holds as soon as $q$
is bounded below on $R_\pi$; a certified model
counter~\cite{kiesl2024certified,chistikov2015approximate} supplies
one by a different route.

Third, the framework prices both routes on one scale.  Enumeration
buys $b_\pi = 0$ at a cost exponential in the region's dimension;
importance sampling buys $b_\pi = O(\sigma_q/\sqrt{m})$ at a cost
linear in $m$.  Both enter the half-width additively, so a
mass-estimation action is comparable to a \textsc{Sample} or
\textsc{Refine} action under the same gain-per-cost rule.
\toolname{} as evaluated never exercises this path, since every
benchmark domain enumerates and $\beta = 0$ throughout; we state the
corollary because it is what the framework claims about the methods it
generalises.

\subsection{The Driver Loop}\label{sec:asip-loop}

\toolname{} is the framework's adaptive instantiation, turning
Theorem~\ref{thm:hw-decomp} into an algorithm by repeatedly choosing
whichever of two actions yields the highest gain per unit cost.
\begin{description}
\item[\textsc{Sample}$(\pi, k)$ (reduces $\epsstat$).]
  Draw $k$ iid samples from $\dist_{\mid R_\pi}$, run $\prog$
  concretely on each sampled input, update the empirical estimates $\hpathat_\pi$ and $W_\pi$. This operation achieves gain $\Delta\epsstat := \hwhat_\pi\,(W_\pi(n_\pi) - W_\pi(n_\pi+k))$ and costs $k$ program evaluations.
\item[\textsc{Refine}$(\pi, b)$ (reduces $\Wopen$).]
  Split $\pi$ along an observed clause $b$ and attempt SMT
  closure on each child node (Lemma~\ref{lem:closure-soundness}). This operation achieves gain  $\Delta\Wopen := \hwhat_\pi\Pr[\text{a child closes}]$ and costs one SMT call per child node.
\end{description}
\paragraph{Drawing from $\dist_{\mid R_\pi}$.}
The cost accounting above charges \textsc{Sample}$(\pi,k)$ exactly $k$
program evaluations, which is only honest if an in-region input can be
obtained in $O(1)$ draws.  It can, and not by rejection.  Let $R_\pi$
be axis-aligned, $R_\pi = \prod_{i=1}^{d}\{\ell_i,\dots,u_i\}$, the
form every refinement along a clause over a single input variable
produces.  Because $\cD$ factors (\Cref{def:d1}), the conditional
$\dist_{\mid R_\pi}$ factors over the same coordinates, and its $i$-th
marginal is the marginal $\cD_i$ truncated to
$\{\ell_i,\dots,u_i\}$.  We therefore sample each coordinate
\emph{directly}, by inverse-CDF on the truncated marginal: draw
$u \sim \mathrm{Unif}[0,1]$, and return the least $x_i$ with
$\cD_i(x_i) \ge \cD_i(\ell_i - 1) + u\,\bigl(\cD_i(u_i) -
\cD_i(\ell_i-1)\bigr)$, found by binary search on the closed-form
CDF.  Every draw lands in $R_\pi$ by construction; none is discarded.
The cost of one in-region input is thus $O(d\log|\inX|)$ arithmetic
operations and exactly one program evaluation, \emph{independent of
$w_\pi$}.

This is the mechanism behind the rare-event results of
\Cref{sec:rq-rare}.  Naive rejection would need $\Theta(1/w_\pi)$
draws from $\cD$ per accepted input, $10^4$ wasted draws on a leaf of
mass $10^{-4}$, and the counts we report would understate the true cost
by that factor; direct conditional sampling removes the factor rather
than hiding it, so the counts in \Cref{sec:results} are total
input-generation cost.  Where $R_\pi$ is \emph{not} axis-aligned, a
clause coupling two input variables, the factorisation fails and the
implementation falls back to rejection or to the importance-sampling
estimator of \Cref{sec:rare-event-instantiation}, at $\Theta(1/w_\pi)$
cost and a $\beta > 0$ mass budget.

Both gains are measured in units of one program execution, and the
loop terminates when $\epsstat + \Wopen \le \varepsilon$ or when a cap
fires (sample budget $B_n$, wall-clock budget $B_t$, or a minimum
gain-per-cost floor).  The loop is then four lines.  Initialise $\partset$ to the single open
leaf $\pi_0$ with $F_{\pi_0} = \top$ and $\hwhat_{\pi_0} = 1$;
bootstrap $n_{\mathrm{boot}}$ samples at $\pi_0$ and attempt closure.
While the stopping rule is unmet, enumerate the candidate
\textsc{Sample} and \textsc{Refine} actions over the open leaves,
execute the one maximising $\mathrm{gain}(a)/\mathrm{cost}(a)$, and
attempt SMT closure on each newly affected leaf.  On exit return
$\muhat$ and $[\hat L,\hat U]$ via \eqref{eq:muhat-def}
and~\eqref{eq:interval}.

Three engineering choices remain open, the per-leaf bound $W_\pi$, the
clause $b$ selected when several are available, and the SMT backend.
By Theorem~\ref{thm:hw-decomp} none affects soundness, only the
constants of $\epsstat$ and the rate at which $\Wopen$ shrinks per SMT
call.

%% file: sections/implementation.tex
\section{\toolname{}: Implementation}\label{sec:implementation}

\toolname{} is an open-source Python prototype that fixes the
three pluggable components of Section~\ref{sec:algorithm} and
the gain-per-cost schedule; by Theorem~\ref{thm:hw-decomp} these
choices affect only $\epsstat$ and $\Wopen$, not soundness.
Mass is computed exactly via marginal-CDF products on axis-aligned
boxes and finite enumeration otherwise, giving $\beta=0$
(Lemma~\ref{lem:mass-conservation}); the certified
importance-sampling estimator of
\Cref{sec:rare-event-instantiation} is retained as a fallback for
non-axis-aligned regions but is invoked by no benchmark in
Section~\ref{sec:results}, so $\beta = 0$ throughout the evaluation
and every reported half-width is $\epsstat + \Wopen$.
Conditional draws use the direct inverse-CDF sampler of
\Cref{sec:asip-loop}: every refinement clause in our benchmarks
constrains a single input variable, so every leaf region is
axis-aligned and no sampled input is ever rejected.
The per-leaf bound is the PrPl-EB confidence sequence of
~\cite{waudbysmith2024betting}, which is
anytime-valid, closed-form, and variance-adaptive; only
anytime-valid bounds satisfy (I3) under a data-dependent stopping
rule.

A leaf is closed only on an \unsat{} verdict for $(F_\pi \land \lnot \mathrm{path}_\pi)$ with all observations agreeing on path
and $\varphi$-value, reducing (I2) to the oracle's soundness
(Lemma~\ref{lem:closure-soundness}).
One further guard is needed: expressions outside LIA (bitwise
operators, shifts) cause the interpreter to emit a
\emph{concretized} constant, producing a trivially-true clause
that would allow vacuous closure; any leaf with a concretized
trace is kept open and contributes to $\Wopen$.
\Apx{app:impl} covers the hybrid interpreter, SMT abstraction and C frontend.

%% file: sections/experiments.tex
\section{Experimental Evaluation}\label{sec:results}

We evaluate \toolname{} through two research questions and compare it
with the two other framework instantiations.  Since all three methods
share the same framework components, the comparison isolates the effect
of the scheduling policy.  The headline result
(\Cref{sec:rq-sched}) is that on the externally curated SV-COMP corpus,
\toolname{}'s adaptive schedule matches or outperforms exhaustive
symbolic refinement on $121$ of $126$ programs.

\subsection{Setup}\label{sec:setup}

\paragraph{Research questions.}
\textbf{RQ1 (Rare events).}
How many program evaluations does \toolname{} require on rare-event
targets compared with plain Monte Carlo?

\textbf{RQ2 (Adaptive scheduling).}
Does the gain-per-cost schedule improve over exhaustive symbolic
refinement under a matched budget?

The coverage guarantee is established theoretically by
Theorem~\ref{thm:hw-decomp}; the experiments measure efficiency and
interval tightness.

\paragraph{Baselines: the three instantiations.}
The comparison is internal to the framework: the three methods differ
only in their scheduling policy.
\code{plain\_mc} is Wilson Monte Carlo, the
\emph{sample-only} schedule;
\code{pse\_exact} is probabilistic symbolic execution in the
Geldenhuys--Dwyer--Visser~\cite{geldenhuys2012pse} /
Filieri~et~al.~\cite{filieri2013spf} lineage, the
\emph{refine-only} schedule;
and \toolname{} is the gain-per-cost schedule using the PrPl-EB
per-leaf bound.
All three share the frontier representation, exact mass estimator, and
closure rule, differing only in how they choose the next action.

\paragraph{Scope of the comparison.}
Rare-event estimators from \Cref{sec:related} are not evaluated as
separate baselines because they address a different component of the
framework.  They provide alternative ways of estimating quantities
required by the invariants rather than alternative scheduling policies.

Moreover, the certification contract requires a finite-sample interval
that remains valid at a data-dependent stopping time.  Many rare-event
methods report a point estimate or an asymptotic confidence interval,
which does not directly satisfy this requirement.  A fair comparison
therefore requires pairing each estimator with an appropriate
time-uniform confidence guarantee before comparing sample efficiency.
Our comparison with \code{plain\_mc} follows this protocol: both methods
produce certified intervals, although \code{plain\_mc} uses a fixed
sample budget while \toolname{} supports adaptive stopping.

\paragraph{Benchmarks.}
We use three benchmark categories selected independently of \toolname{}.
\emph{Category A} contains twelve introductory and textbook integer
kernels (Hacker's Delight~\cite{warren2012hd},
CLRS~\cite{cormen2009clrs}, Knuth TAoCP~\cite{knuth1998taocp},
a CERT-C overflow check, and Collatz~\cite{lagarias2010collatz}),
together with \code{dispatch\_table}, a path-explosive kernel.

\emph{Category B} is drawn from the SV-COMP ReachSafety-Loops corpus
\cite{beyer2024svcomp}, processed using the C frontend described in
\apx{sec:c-frontend}.

\emph{Category C} contains four integer kernels taken from production
code: FreeRTOS \code{queue.c}, lwIP \code{inet\_chksum.c},
zlib \code{crc32.c}, and SQLite \code{util.c}.
\Apx{app:benchmarks} lists each benchmark, its property, and exact
$\mutrue$.

\paragraph{From the SV-COMP corpus to Category B.}
SV-COMP programs do not provide operational input distributions.  We
therefore model each input using
$\mathrm{BG}(p{=}0.1,N{=}100)$, a bounded geometric distribution that
assigns higher probability to smaller integers.

The original $601$-program corpus is reduced to the evaluated subset
using four structural criteria, independent of any method's output:
$158$ programs fall outside the supported integer fragment (pointers,
floats, escaping calls), $86$ have no
\code{\_\_VERIFIER\_nondet} input and therefore constant reliability,
$84$ raise runtime errors under concolic tracing, and $147$ exceed the
per-program wall-clock limit because of divergent loops.  This leaves
$126$ programs evaluated end-to-end.

\paragraph{Protocol.}
Unless noted otherwise, each experiment uses sample budget $2000$,
$\delta=0.05$, three random seeds, the Z3 backend
\cite{demoura2008z3}, and a per-cell wall-clock limit.  We report
medians.  Each input domain is finite, so $\mutrue$ is well defined
and, where feasible, computed exactly by enumeration.

\paragraph{Soundness.}
Theorem~\ref{thm:hw-decomp} establishes the coverage guarantee for any
instantiation satisfying the three invariants.  \toolname{} satisfies
these invariants (\Cref{sec:implementation}), so the experiments focus
on efficiency and interval tightness.  The scheduler determines which
actions are attempted, but not whether a refinement or closure result is
sound.

\subsection{RQ1: Sample Efficiency on Rare Events}\label{sec:rq-rare}

Plain Monte Carlo requires
$\Theta(1/(\mutrue\varepsilon^2))$ samples to certify a rare event
($\mutrue\ll1$) to half-width $\varepsilon$.
In contrast, once \toolname{} closes the dominant regions of the input
space, the remaining certification cost can become largely independent
of $\mutrue$.

\begin{figure}[t]
  \centering
  \includegraphics[width=0.48\textwidth]{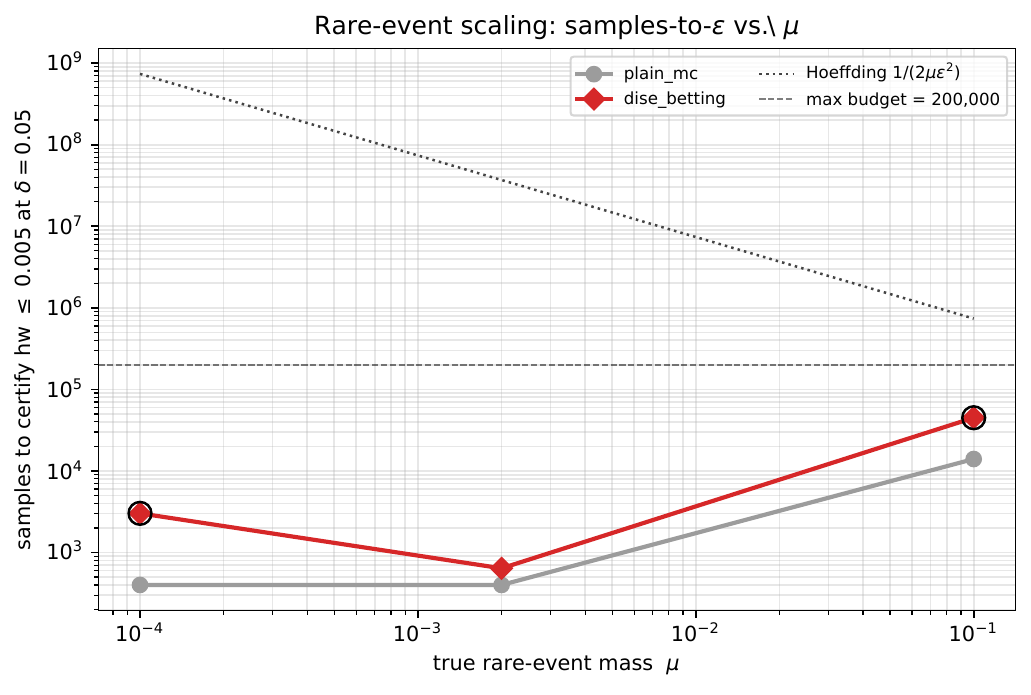}
  \hfill
  \includegraphics[width=0.48\textwidth]{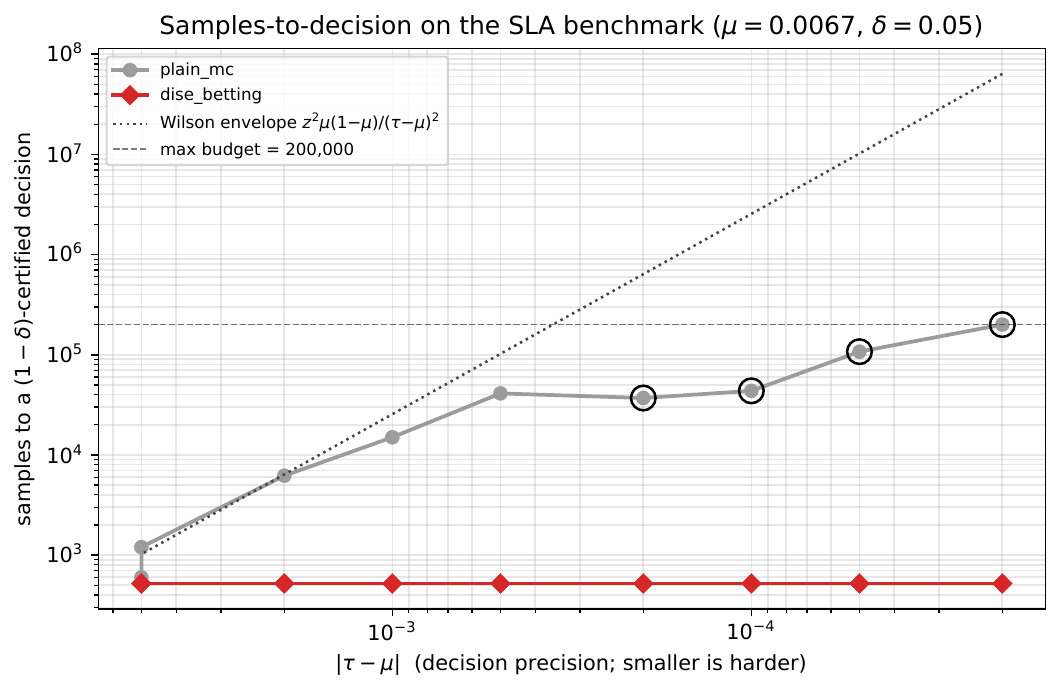}
  \caption{RQ1.  \emph{Left:} samples to certify $\mutrue$ to half-width
    $5\times10^{-3}$ against $\mutrue$ (log-log); the dotted line is the
    Hoeffding envelope for plain Monte Carlo.  \emph{Right:} samples to a
    $(1-\delta)$-certified SLA decision against decision precision
    $|\tau-\mutrue|$ (log scale).  Budget-capped points are hollow.}
  \label{fig:rq1}
\end{figure}

On an SLA-style load classifier
$[\mathrm{load}\ge5000]$ under
$\mathrm{BG}(10^{-3},10^4)$ inputs
($\mutrue\approx6.68\times10^{-3}$), we vary the decision precision
$|\tau-\mutrue|$ (\Cref{fig:rq1}, right).
\toolname{} reaches a certified decision in approximately $520$ samples
across five orders of magnitude of precision, while plain Monte Carlo
follows the Wilson envelope and does not decide within
$2\times10^5$ samples once
$|\tau-\mutrue|\le5\times10^{-4}$.

On a parametric rare-event benchmark with
$\mutrue=w/N$ swept across four orders of magnitude
(\Cref{fig:rq1}, left), \toolname{} certifies to
$\varepsilon=5\times10^{-3}$ in $\Theta(1)$ samples for
$\mutrue\in[5\times10^{-4},2\times10^{-3}]$,
while plain Monte Carlo follows the expected rare-event scaling.
For larger probabilities, above approximately
$\mutrue\approx10^{-2}$, plain Monte Carlo becomes more efficient
because \toolname{} incurs the additional cost of SMT refinement.

\subsection{RQ2: Adaptive Scheduling vs.\ Exhaustive
Refinement}\label{sec:rq-sched}

\code{pse\_exact} and \toolname{} share the frontier, mass estimator,
and closure rule, differing only in scheduling policy.  We therefore use
their difference to evaluate the effect of adaptive scheduling.

We evaluate all $126$ Category-B programs from SV-COMP
(Table~\ref{tab:svcomp}), a benchmark set curated independently of
\toolname{}.  Under the
$\mathrm{BG}(0.1,100)$ input model, probability mass is concentrated on
smaller inputs, and the adaptive scheduler prioritises the corresponding
regions.

\toolname{} produces an exact ($\mathrm{hw}=0$) interval on $59$ of the
$126$ programs, compared with $51$ for exhaustive
\code{pse\_exact}.  Its median half-width is $0.0030$, compared with
$0.043$ for \code{pse\_exact}.  Across individual programs,
\toolname{} is strictly tighter on $50$ programs, ties on $71$, and is
outperformed on $5$.  Thus, the adaptive schedule matches or improves
upon exhaustive refinement on $121$ of $126$ programs.

Compared with the sample-only schedule, \code{plain\_mc} wins on
$62$ programs and \toolname{} wins on $64$.  \code{plain\_mc} has a
smaller median half-width because Wilson intervals are narrow on many
near-zero-$\mutrue$ instances, whereas \toolname{} can additionally
produce exact certificates through symbolic closure.

\begin{table}[t]
  \centering
  \caption{RQ2 at scale: the three schedules on the $126$ Category-B
    (SV-COMP) programs under $\mathrm{BG}(0.1,100)$ inputs.
    ``Exact'' counts $\mathrm{hw}=0$ certificates.}
  \label{tab:svcomp}
  \small
  \begin{tabular}{lrr}
    \toprule
    Method & Median half-width & Exact ($\mathrm{hw}=0$) \\
    \midrule
    \code{plain\_mc}  & $0.0005$          & $0/126$ \\
    \code{pse\_exact} & $0.043$           & $51/126$ \\
    \toolname{}       & $\mathbf{0.0030}$ & $\mathbf{59/126}$ \\
    \bottomrule
  \end{tabular}
\end{table}

\subsection{A Certification Threshold: Sample Budget}
\label{sec:case-study}

We compare the number of program evaluations required to certify an
extremely high reliability target.  This analysis is analytical rather
than experimental and is based on the confidence bounds used by the
different certification strategies.

Let the certification threshold be $\tau=10^{-9}$ and the confidence
parameter be $\delta=0.05$.  The goal is to certify
$\mu\ge1-\tau$, equivalently that the failure probability
$1-\mu$ is at most $\tau$.

\paragraph{Sampling-only certification.}
Suppose $P$ is evaluated on $n$ iid samples from $\mathcal D$ and no
failures are observed.  The $(1-\delta)$ upper confidence bound on the
failure probability is approximately
$\ln(1/\delta)/n$, which equals $3/n$ for $\delta=0.05$.
Therefore,
\[
  \frac{3}{n}\le\tau
  \qquad\Longrightarrow\qquad
  n\ge\frac{3}{\tau}=3\times10^9 .
\]

Thus, even under the favourable assumption of observing no failures,
sampling alone requires roughly three billion program executions.

\paragraph{Certification with symbolic closure.}
Suppose symbolic reasoning proves all but a residual region $R$ of mass
$w$, so that the total failure probability is
\[
1-\mu=w\,p_R ,
\]
where $p_R$ is the conditional failure probability inside $R$.

If $w\le\tau$, the residual mass alone is below the certification
threshold and no additional sampling is required.  Otherwise, sampling
is performed only within $R$, giving
\[
  \frac{3}{n_R}\le\frac{\tau}{w}
  \qquad\Longrightarrow\qquad
  n_R\ge\frac{3w}{\tau}.
\]

Compared with sampling over the original distribution, the required
sample budget is reduced by a factor of $1/w$.  For example,
$w=10^{-3}$ reduces the budget from $3\times10^9$ to
$3\times10^6$ evaluations, while $w=10^{-6}$ reduces it to
$3\times10^3$.

\paragraph{Anytime-valid confidence bounds.}
\toolname{} uses an anytime-valid confidence sequence rather than a
fixed-$n$ interval.  The additional width introduced by the
time-uniform guarantee changes the constants but not the asymptotic
scaling of the sample requirements.  In particular, the logarithmic
factors from confidence sequences and the Bonferroni correction in
(I3) increase the required budget by only a small multiplicative factor.

The analysis assumes that the operational distribution $\mathcal D$ is
known exactly.  When $\mathcal D$ is estimated from data, uncertainty in
the distribution must also be incorporated into the reliability
guarantee; extending the framework to distributionally robust
certification is left for future work.

%% file: sections/conclusion.tex
\section{Conclusion}\label{sec:conclusion}

We presented distribution-aware reliability estimation through a decomposition
of the certification error into two components: a statistical error
$\epsstat$, reduced by sampling, and a structural error $\Wopen$, reduced by
symbolic refinement. Theorem~\ref{thm:hw-decomp} shows that any combination of
a mass estimator, a closure rule, and a per-leaf confidence bound satisfying
the invariants of Definition~\ref{def:invariants} yields a
$(1-\delta)$-coverage reliability interval. Pure Monte Carlo, probabilistic
symbolic execution, and the rare-event estimators of
Corollary~\ref{cor:certified-is} arise as different instantiations of these
components rather than different soundness arguments.

\paragraph{Limitations and future work.}
Most limitations arise from the current implementation rather than from the
framework itself. The assumptions of determinism
(\Cref{ass:program}), discrete product input distributions
(\Cref{def:d1}), Boolean output properties
(\Cref{def:bool-property}), and linear integer arithmetic simplify the current
instantiation but are not required by
Theorem~\ref{thm:hw-decomp}. Extending the framework therefore amounts to
replacing individual components while preserving the invariants.

\begin{enumerate}
    \item \emph{Distribution class.}
    Definition~\ref{def:d1} assumes independent input coordinates.
    Supporting Bayesian-network or other dependent discrete distributions
    requires a mass estimator and conditional sampler that account for
    dependencies, replacing the inverse-CDF construction of
    \Cref{sec:asip-loop}.

    \item \emph{Determinism and nondeterminism.}
    Lemma~\ref{lem:closure-soundness} relies on determinism to certify the
    property value of an entire region from a single execution path.
    Randomized programs fit the framework by exposing the random seed as part of
    the input domain, as assumed in \Cref{ass:program}. Adversarial
    nondeterminism, following Luckow et al.~\cite{luckow2014exact}, requires
    per-leaf bounds over all resolutions and remains future work.

    \item \emph{Property class.}
    We consider Boolean predicates over program outputs
    (\Cref{def:bool-property}). Temporal properties such as LTL can be handled
    by composing the program with a monitor automaton, reducing monitor
    acceptance to a Boolean output property, but constructing this product is
    outside the scope of the present implementation.

    \item \emph{Bitvector semantics.}
    The current solver targets linear integer arithmetic. Traces involving
    unsupported bitvector operations are conservatively left open, increasing
    $\Wopen$ without affecting soundness. A \code{QF\_BV} backend would recover
    machine-word semantics at the cost of more expensive SMT solving.

    \item \emph{Refinement strategy.}
    Refinement currently uses predicates observed along explored execution
    paths. Predicate synthesis techniques such as Craig interpolation provide a
    natural extension.

    \item \emph{Uncertain operational distributions.}
    The certification guarantee is conditional on the input distribution
    $\mathcal D$. When $\mathcal D$ is estimated from data, uncertainty in the
    distribution propagates to the reliability estimate
    (\Cref{sec:case-study}). Extending the framework to distributionally robust
    certification is an important direction for future work.
\end{enumerate}

These extensions modify the implementation, not the soundness theorem.
Similarly, adversarial nondeterminism~\cite{luckow2014exact} and temporal
specifications extend the class of certifiable systems without changing the
underlying certification contract. Replacing importance sampling with certified
approximate model counting
~\cite{kiesl2024certified,yang2023approxmc6,chistikov2015approximate}
is another promising direction toward end-to-end machine-checkable reliability
certificates.

\paragraph{Reproducibility.}
The implementation of \toolname{}, benchmark suite, frontend, and scripts used
to regenerate all tables and figures are archived under the MIT licence at
\url{https://doi.org/10.5281/zenodo.21835008}. The extended version contains
the complete proofs, and \code{scripts/reproduce.sh} reproduces the experimental
results (\code{QUICK=1} runs a smoke test).

\subsection*{Acknowledgement}

This work was supported by the Anusandhan National Research Foundation, Government of
India, under the Prime Minister Early Career Research Grant ANRF/ECRG/2025/001136/ENS, and by the Vachani School of Advanced Computing, Ashoka University under the Young Faculty Grant (00281). 